\def\breakon{\end{multicols}\widetext\vspace{.5cm}
\noindent\rule{.48\linewidth}{.3mm}\rule{.3mm}{.5cm}\vspace{.5cm}}
\def\breakoff{\vspace{.5cm}
\noindent
\rule{.52\linewidth}{.0mm}\rule[-.47cm]{.3mm}{.5cm}\rule{.48\linewidth}{.3mm}
\vspace{.5cm}
\begin{multicols}{2}
\narrowtext}
\documentstyle[aps,epsf,multicol,eqsecnum]{revtex}
\begin{document}
\draft
\widetext
\title{What do noise measurements reveal about fractional
charge in FQH liquids?}
\author{Nancy P.\ Sandler, Claudio de C.\ Chamon and Eduardo Fradkin}
\address{Department of Physics, University of Illinois at Urbana-Champaign,
Urbana, IL 61801-3080}
\maketitle
\begin{abstract}
We present a calculation of noise in the tunneling current through
junctions between two two-dimensional electron gases (2DEG) in
inequivalent Laughlin fractional quantum Hall (FQH) states, as a
function of voltage and temperature. We discuss the interpretation of
measurements of suppressed shot noise levels of tunneling currents
through a quantum point contact (QPC) in terms of tunneling of
fractionally charged states. We show that although this interpretation
is always possible, for junctions between different FQH states the
fractionally charged states involved in the tunneling process are not
the Laughlin quasiparticles of the isolated FQH states that make up
the junction, and should be regarded instead as solitons of the
coupled system. The charge of the soliton is, in units of the electron
charge, the harmonic average of the filling fractions of the
individual Laughlin states, which also coincides with the saturation
value of the differential conductance of the QPC. For the especially
interesting case of a QPC between states at filling fractions $\nu=1$
and $\nu={\frac{1}{3}}$, we calculate the noise in the tunneling
current exactly for all voltages and temperatures and investigate the
crossovers. These results can be tested by noise experiments on
$(1,{\frac{1}{3}})$ QPCs.  We present a generalization of these
results for QPC's of arbitrary Laughlin fractions in their weak and
strong coupling regimes. We also introduce generalized Wilson ratios
for the noise in the shot and thermal limits. These ratios are
universal scaling functions of $V/T$ that can be measured
experimentally in a general QPC geometry.
\end{abstract}

\pacs{PACS: 73.40.Hm, 71.10.Pm, 73.40.Gk, 73.23.-b}

\begin{multicols}{2}
\narrowtext
\section{INTRODUCTION}
\label{sec:intro}

\begin{figure}
\vspace{0.2cm}
\hspace{.5in}
\epsfxsize=2in
\epsfbox{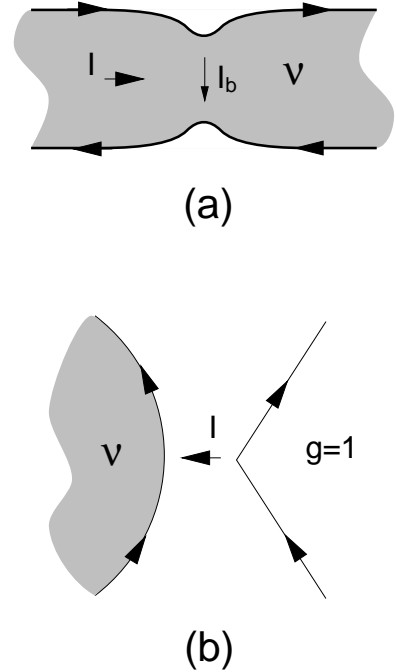}
\vspace{0.9cm}
\noindent
\caption{Different geometries for measuring shot noise in FQH systems:
a) Geometry with tunneling between the edges of the same FQH systems
with filling fraction $\nu$; b) Geometry with tunneling between the
edges of two different FQH system with filling fractions $\nu$ and 1
(representing a normal metal) respectively. }
\label{fig:fig-junctions}
\end{figure}

Recently, two experimental groups, in Saclay \cite{Glattli} and at the
Weizmann Institute \cite{Reznikov}, have been able to measure
suppressed shot noise in a quantum point contact (QPC) geometry - a
constriction in the plane of a 2DEG. In this setup the two edges of
the FQH system are brought together by applying a gate voltage that
creates the QPC. In what follows we will refer to this particular
geometry as tunneling between the edges of the {\it same} FQH system
(see Fig.~\ref{fig:fig-junctions}(a)). The quantum shot noise in this case
reflects the fluctuations in the tunneling current that result from
the presence of the constriction. The results obtained in these
experiments for FQH systems at filling fraction $\nu = 1/3$ are
consistent with the interpretation of uncorrelated tunneling events of
fractionally charged quasiparticles ($e^*=e/3$) between the edges of
the FQH system at the QPC.

At filling fraction $\nu={\frac{1}{3}}$, the Hall conductance
and the fractional charge of the quasiparticles are determined by the
same universal coefficient, the filling fraction. Thus, it is natural
to ask if the noise experiments measure the fractional charge or the
conductance.  Clearly, one way to address this issue is to extend
these measurements to a range of filling fractions not in the Laughlin
sequence, where the charges of the quasiparticles (in units of $e$)
are not equal to the filling fraction. However, the theory of
tunneling in generic FQH states is only understood qualitatively and
many important issues, such as edge reconstructions, still need to be
understood. In contrast, there exists a rather detailed and well
understood theory of tunneling between the edges of the Laughlin
states.
Shot noise measurements can be an important probe of the edges of FQH
states in other, more general, geometries. From this point of view, we
consider the problem of tunneling from a Fermi liquid ({\it i.\ e.\/},
$\nu=1$) to a Laughlin FQH state at filling fraction $\nu$ or, more
generally, between FQH edge states with {\it different} filling
fractions, $(\nu_1,\nu_2)$. An instructive example is presented in
Fig.~\ref{fig:fig-junctions}(b) which is a schematic representation inspired by
the geometry used in the experiments by A.\ Chang and coworkers
\cite{Chang,Matt}.

\begin{figure}
\vspace{0.5cm}
\hspace{.0in}
\epsfxsize=3.0in
\epsfbox{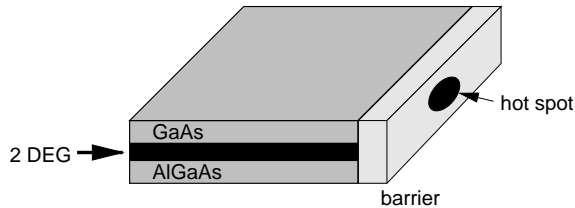}
\vspace{0.9cm}
\noindent
\caption{A possible experimental realization of the
QPC junction between an electron gas and a FQH liquid, inspired by the
cleaved edge overgrowth samples of A.\ Chang and coworkers. The
electron gas reservoirs make contact to the FQH liquid formed in the
2DEG via tunneling through a ``hot spot'', where the barrier is made
weaker. This tunneling region should be small, of the order of a few
cyclotron lengths.}
\label{fig:fig-hotspot}
\end{figure}

A possible experimental realization of the geometry depicted in
Fig.~\ref{fig:fig-junctions} (b) is suggested in Fig.~\ref{fig:fig-hotspot},
where the strong tunneling region or ``hot spot'' is assumed to be
designed to be small, of the order of a few cyclotron lengths. These
sizes are similar to the ones in the QPC experiments of
Refs. \cite{Glattli,Reznikov}.

\begin{figure}
\vspace{0.4cm}
\hspace{.5in}
\epsfxsize=2.0in
\epsfbox{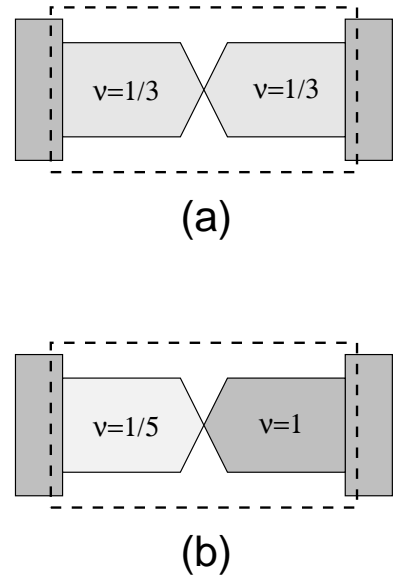}
\vspace{0.9cm}
\noindent
\caption{Two different junctions $(\nu_1,\nu_2)$ which cannot be told
apart if one only looks at the two-terminal properties measured
outside of the dashed region. The effective Luttinger parameter
$g=\frac{2\nu_1\nu_2}{\nu_1+\nu_2}$ is the same in both
cases. Therefore, the $(\frac{1}{3},\frac{1}{3})$ junction in (a) and
the $(\frac{1}{5},1)$ junction in (b) have both: {\it i}) the same
asymptotic conductance at large voltages $G=\frac{1}{3}\frac{e^2}{h}$,
{\it ii}) the same shot noise level $S=2eI$ for small voltages, and
{\it iii}) the same shot noise level $S=2\frac{e}{3}I_b$ for small
backscattering currents $I_b= GV-I$}
\label{fig:fig-nancyparadox}
\end{figure}

In this paper we consider the more general problem of non-equilibrium
noise in tunnel junctions between different FQH states. We will show
here that, while noise measurements always constitute a detailed probe
of the theory of the edge states, their interpretation in terms of the
(generally fractional) charges of the quasiparticles is actually quite
subtle. In particular, it is both natural and important to inquire if
measurements of suppressed shot noise level in a generic QPC junction
provide unambiguous evidence for fractional charge in the isolated FQH
fluid and what is its relation with the conductance. In the example
that we will discuss here, we will find that the level of the shot
noise in the weak backscattering regime tracks the differential
conductance of the junction instead of being determined by the
(fractional) charge of the FQH quasiparticles or by the electron
charge. In fact, we will also discuss an example in which in a
two-terminal measurement of the shot noise level, it is impossible to
distinguish between tunneling of electrons between the edges of
identical Laughlin states and electrons between (carefully chosen)
different FQH states (see Fig.~\ref{fig:fig-nancyparadox}), for
instance $({\frac{1}{3}},{\frac{1}{3}})$ and $({\frac{1}{5}},1)$.
Logically, there are two related issues involved in this problem. One
is the charge of the quasiparticle participating in the tunneling
process, and another are the properties of the isolated fluids
connected by the junction.

Let us begin by reviewing the basic theoretical assumptions and
interpretations involved in the various studies of noise in tunneling
between Luttinger liquids \cite{K&F-noise,fls,FS2,LS3,CFW1,CFW}. Within the
theoretical framework of tunneling between the edges of a given FQH
system, the low temperature (shot) noise spectrum $S$ is calculated in
terms of the correlation function of the tunneling current, which in
the geometry of Fig.~\ref{fig:fig-junctions} (a)  corresponds to the
backscattering current $I_b$. In this limit, the shot noise level $S$
is
\begin{equation}
S= 2 q I_b
\label{eq:noise}
\end{equation}
This result can be interpreted as meaning that the backscattering
current is due to `uncorrelated' tunneling of Laughlin quasiparticles
carrying fractional charge, {\it i.e.}, the current corresponds to a
sequence of uncorrelated Poisson distributed quasiparticles that
tunnel through the QPC. This result applies provided the
backscattering current $I_b$ is arbitrarily small.  Thus, the shot
noise level for small tunneling currents reflects the charge of the
carriers and, consequently, in the geometry of
Fig.~\ref{fig:fig-junctions} (a), it measures the fractional charge.
Alternatively, in the same geometry, we may also regard the
transmission current as being carried by electrons, which are strongly
correlated. This description corresponds to the dual picture of
Ref.~\cite{KF-dual}. In this picture, defects in the transmission of
electrons correspond to backscattering of kinks ({\it i.\ e.\/},
``magnetic charges'') or quasiparticles.  Thus, in this
representation, the suppression factor in the shot noise level is a
measure of the correlations among the electrons. In this dual picture,
although the transmission current is {\it large}, the fluctuations of
this current ({\it i.\ e.\/}, the noise) are small and due to defects
(or kinks) which play the role of the backscattering current $I_b$ in
the quasiparticle picture.

Naturally, both pictures are completely equivalent and consistent with
each other, and they yield the same result, as they
should. Nevertheless, it is worth to stress that the physical
interpretation of the coefficient $q$ of the shot noise as the charge
of the quasiparticle of the FQH state is not based on a direct
measurement of the charge.  Instead, this interpretation relies on the
existence of the quasiparticle picture since it is precisely the
tunneling of quasiparticles which causes the fluctuations of the
current in that picture. This interpretation is physically consistent
because it is possible to determine independently by a transport
measurement that the 2DEG is in a Laughlin state.

From a conceptual point of view it is then natural to ask if it is
always possible to find an analog of the quasiparticle picture in
which the coefficient of the shot noise is always necessarily
determined by the charge of a physical eigenstate of an {\it isolated}
FQH state. We will see now that answering this question leads to an
interesting paradox, shown in Fig.~\ref{fig:fig-nancyparadox}. Let us
reexamine the problem of quasiparticle tunneling between the edges of
a $\nu={\frac{1}{3}}$ Laughlin state but as seen from the dual
picture. In this representation, we have {\it electrons} tunneling
{\it between} two $\nu={\frac{1}{3}}$ Laughlin states. It turns out
that, using the methods of ref.\
\cite{cl+ed}, it is straightforward to show (see below) that this
tunnel junction is equivalent to a tunnel junction between Laughlin
states $\nu=1$ and $\nu={\frac{1}{5}}$ in which {\it electrons} hop
between the two fluids at the QPC. The paradox resides in the fact
that the noise in the tunneling current is determined by a coefficient
which is still $q={\frac{1}{3}}$ in spite of the fact that there are
no such quasiparticles in the {\it bulk} of the isolated $\nu=1$ and
$\nu={\frac{1}{5}}$ states. In this picture, $q={\frac{1}{3}}$ is the
charge of the soliton which diagonalizes the junction
Hamiltonian. From the point of view of a $(1,{\frac{1}{5}})$ junction,
these states are complicated non-local superpositions of the quantum
states of the two isolated inequivalent FQH fluids. The only way to
distinguish cases (a) and (b) in Fig.~\ref{fig:fig-nancyparadox} is to
determine the conductance of the fluids independently through a
four-terminal measurement. In fact, a two-terminal measurement cannot
distinguish these two cases. It is also worth to remark that, as we
will see below, the coefficient of the noise level of picture (b) in
Fig.~\ref{fig:fig-nancyparadox}, is precisely the same (in units of
${\frac{e^2}{h}}$) as the saturation value of the differential
conductance of the $(1,{\frac{1}{5}})$ junction!

One may also ask, given a QPC between two inequivalent FQH states, if
it is always possible to find an equivalent system in which the noise
in the tunneling current can be interpreted as due to tunneling of
quasiparticles of isolated FQH fluids. The answer to this question is
{\it no}.  We will see below that a generic QPC $(\nu_1,\nu_2)$, for
the Laughlin states $\nu_i^{-1}=2k_i+1$ and $k_i\in {\bf Z}$
($i=1,2$), is equivalent to the QPC $(\nu_{eff},\nu_{eff})$ with
$\nu_{eff}^{-1}=k_1+k_2+1$. Hence, the equivalent junction represents
tunneling between {\it fermion} Laughlin states only if either $k_1$
or $k_2$ (but not both) are odd integers. However, what it is always
true, is the statement that the noise level is determined by the
charge of the soliton (kinks) states that diagonalize the Hamiltonian
of the junction and, in general, these states cannot be represented
simply in terms of the quasiparticles of the isolated
fluids. Superficially, this result may appear to violate the basic
bulk-edge correspondence which is crucial for the theory of edge
states of FQH fluids\cite{wen-edge}, as it involves tunneling of
objects which are neither electrons nor quasiparticles of the isolated
fluids.  Actually, as we discuss below (see also the appendix of
Ref.~\cite{andreev}), this is not a real paradox or contradiction
since the edges are gapless and the structure of the Hilbert space is
respected by these tunneling processes.

Measurements of noise, in addition to being useful tools to
investigate experimentally the problem of fractional charge in FQH
fluids, can also be used to probe the properties of the edges in
greater detail.  For instance, for the special and exactly solvable
case of the $(1,{\frac{1}{3}})$ junction, we calculate the current
correlation functions exactly for all voltages, temperatures and
tunneling amplitudes, and show that the temperature and voltage
dependence of the noise contains a great deal of information on the
edge states, on the fixed points of the junctions and of their
crossovers. For a generic $(\nu_1,\nu_2)$ junction, for which the
correlation functions cannot be computed exactly, we find the noise in
the asymptotic regimes of large and small voltages, and for high and
low temperatures. The results are discussed in the form of a phase (or
rather, crossover) diagram. To analyze the information obtained from
the asymptotic regimes, we introduce a generalization of Wilson ratios
useful in quantum impurity problems
\cite{andrei}. We define a generalized Wilson ratio as the quotient
between the shot noise and the thermal noise levels. Near the two
fixed points this ratio becomes a universal scaling function of $V/T$
independent of the coupling constant. The ratio contains information
on the Luttinger liquid behavior through the parameter $g'$ both in
the exponent of the $V/T$ dependence and in the constant $g'$
dependent prefactor. The ratios in the two fixed points are also
related by duality. Finally, these ratios can be used to analyze the
experimental data from both limits of thermal and shot noise in a
unified way.

The paper is organized as follows. In Section~\ref{sec:model} we
review the model of Refs.\ \cite{cl+ed,andreev} and we introduce the
relations among the charge densities of the rotated and the dual
fields. We also define the backscattering current $I_b$ and introduce
the definitions for the noise in both currents $S_I$ and $S_{I_b}$.
In section~\ref{sec:mismatch} we discuss a generic junction
between two FQH states at filling fractions $\nu_1$ and $\nu_2$. Here
we give a general result for the current shot noise at both small and
large tunneling amplitudes. In Section
\ref{sec:noiset0} we consider in detail the $(1,{\frac{1}{3}})$
junction. Here we review (briefly) the refermionization procedure used
to diagonalize exactly the Hamiltonian for this junction
\cite{CFW}. For this particular case, we calculate the noise at zero
temperature in the current through the junction. We show that
the noise in the limit of strong coupling (or weak backscattering
current) is not a direct measurement of the fractional charge of the
decoupled FQH fluid but, instead, it measures the charge of the
soliton that diagonalizes the Hamiltonian, which also determines the
saturation value of the conductance of the junction. In Section
\ref{sec:noiset} we calculate the noise in the current $I$ and
the backscattering current $I_b$ as a function of temperature $T$ and
voltage $V$. We present our results in the form of a $T-V$ diagram for
the noise. In particular, we show that the strong coupling regime
relates the equilibrium and non-equilibrium regimes of the
junction. The weak coupling region characterizes the regime of low
temperatures and low voltages. We introduce ratios between the thermal
and shot noise limits for both currents $I$ and $I_b$.  In section
\ref{sec:ratios} we define generalized Wilson ratios as quotients 
between noise in the thermal and shot noise regimes for junctions with
a generic value of $g'$ using perturbative methods. These ratios are
universal scaling functions of $V/T$ around both the weak and the
strong coupling fixed points. In particular, we calculate the value of
these ratios for the geometry depicted in
Fig.~\ref{fig:fig-junctions} (a) used in recent experiments by
L.~Saminadayar et.~al\cite{Glattli} and R.~de~Picciotto\cite{Reznikov}
for a $(\frac{1}{3}, \frac{1}{3})$ junction.
In section \ref{sec:auto-correlations}
we discuss a four probe geometry and calculate the noise in all four
channels. We show that, as a consequence of chirality, the noise in
the incoming channels is insensitive to the presence of the QPC; it is
given by the Johnson-Nyquist noise level, and it is proportional to
the conductance determined by the respective filling fraction in
either side of the junction. The noise in the outgoing channels, on
the other hand, contains the information on the tunneling coupling,
and depends on both filling fractions $\nu_1$ and $\nu_2$. Finally, in
Section\ref{sec:conclusions} we summarize our main results and discuss
its experimental implications.

\section{Model for the junction}
\label{sec:model}

In this section we review the model for a mismatched FQH junction used
in Ref.\cite{andreev}. We start with a Lagrangian for the FQH-normal
metal junction that describes the dynamics on the edge of a FQH
liquid, the electron gas reservoirs, and the tunneling between them
through a single point-contact of the form
\begin{equation}
{\cal L} = {\cal L}_{edge} + {\cal L}_{res} + {\cal L}_{tun}\,.
\label{eq:Lag}
\end{equation}
The dynamics of the edge of the FQH liquid with a Laughlin filling
fraction $\nu=\frac{1}{2k+1}$ is described by a free chiral boson
field $\phi_1$ with the Lagrangian \cite{XGWcll}
\begin{equation}
{\cal L}_{edge} = \frac{1}{4\pi}
\partial_x \phi_1 (\partial_t - \partial_x) \phi_1\,.
\label{eq:Led}
\end{equation}
The edge electron and quasiparticle operators are given by
\begin{equation}
\psi_{e} \propto \,:e^{-i\frac{1}{\sqrt{\nu}} \phi_1(x,t)}:\,\,\,;\,\,\,
\psi_{qp} \propto \,:e^{-i \sqrt{\nu} \phi_1(x,t)}:
\label{eq:psiedge}
\end{equation}

${\cal L}_{res}$ describes the dynamics of the electron gas
reservoir. As shown in Ref. \cite{cl+ed}, a 2D or 3D electron gas can
be mapped to a 1D chiral Fermi liquid (FL) ($\nu = 1$) when the
tunneling is through a single point-contact. This 1D chiral Fermi
liquid is represented by a free chiral boson field $\phi_2$. ${\cal
L}_{res}$ is given by
\begin{equation}
{\cal L}_{res} = \frac{1}{4\pi} \partial_x \phi_2 (\partial_t - \partial_x)
\phi_2\,.
\label{eq:Lres}
\end{equation}
In this case, the electron operator is given by
\begin{equation}
\psi_{res} \propto :e^{-i \phi_2(x,t)}:
\label{eq:el}
\end{equation}
The tunneling Lagrangian between the FQH system and the reservoir is
\begin{equation}
{\cal L}_{tun} = \Gamma \; \delta (x)\; e^{-i \omega_0 t}
\; :e^{i[\frac{1}{\sqrt{\nu}} \phi_1(x,t) - \phi_2(x,t)]}:
+{\rm h.\ c.\ }\,,
\label{eq:Ltun}
\end{equation}
where $\Gamma$ represents the strength of the electron tunneling
amplitude which takes place at a single point in space $x=0$, the QPC. 
In what follows, by analogy with quantum impurity problems,
we will refer to the QPC as the impurity.

The voltage difference between the two sides of the junction is introduced
in the model by letting $\Gamma \rightarrow \Gamma e^{-i
\omega_0 t}$, where $\omega_0 = e V/\hbar$. The external
voltage $V$ can be interpreted as the difference between the chemical
potentials of the two systems: $V = \mu_1 - \mu_{\nu}$. 

\begin{figure}
\vspace{.2cm}
\noindent
\hspace{.05in}
\epsfxsize=3.0in
\epsfbox{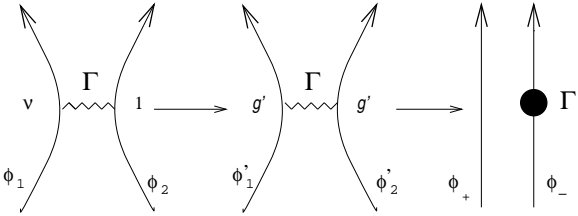}
\vspace{0.9cm}
\caption{Set of rotations from $\phi_{1,2}$ to $\phi_{\pm}$. The first
rotation maps the original junction between two different FQH liquids
to one between FQH with the same filling fraction $g'$, the second one
decouples the problem into two separate ones.}
\label{fig:fig-rotations}
\end{figure}

The density operators at both sides of the junction are defined
as follows:
\begin{equation}
\rho_1 = \frac{\sqrt{\nu}}{2\pi} \partial_x \phi_1 \,\; ,\;\;\;\;\;
\rho_2 = \frac{1}{2\pi} \partial_x \phi_2
\label{eq:densities}
\end{equation}
By a suitable rotation the original Lagrangian ${\cal L}$ can be
mapped into a new one \cite{cl+ed,andreev}:
\begin{eqnarray}
{\cal L}& = & \frac{1}{4\pi}
\partial_x \phi'_1(\partial_t - \partial_x)\phi'_1 +
\frac{1}{4\pi}
\partial_x \phi'_2(\partial_t - \partial_x)\phi'_2 \nonumber\\
        &\ & \ +
\Gamma\, \delta(x)\, e^{-i \omega_0 t}
e^{i \frac{1}{\sqrt{g'}}[\phi'_1(x,t) - \phi'_2(x,t)]} + h.c.
\label{eq:Lab}
\end{eqnarray}
where the new fields $\phi'_1$ and $\phi'_2$ have been introduced, and
$g'$ is an effective Luttinger parameter (which can also be regarded as an
``effective filling fraction") given by:
\begin{equation}
g'^{-1} = \frac{(1+\nu^{-1})}{2}.
\label{eq:g'}
\end{equation}
In particular, this rotation relates the {\it densities} as follows:
\begin{equation}
\left(\matrix{\rho_1 \cr \rho_2}\right)
=
\left(\matrix{
\frac{1 + \sqrt{\nu}}{2}    & \frac{\sqrt{\nu} - 1}{2}\cr
\frac{1 - \sqrt{\nu}}{2\sqrt{\nu}} & \frac{1 + \sqrt{\nu}}{2\sqrt{\nu}}}\right)
\left(\matrix{
\rho'_{1}\cr \rho'_{2} } \right)
\label{eq:rot1}
\end{equation}
Next, we introduce the fields $\phi_-$ and $\phi_+$  that separate
${\cal L}$ into two decoupled Lagrangians ${\cal L}_+$ and ${\cal L}_-$,
\begin{equation}
\phi_+ = \frac{\phi'_1 + \phi'_2}{\sqrt{2}}\,\; ,\;
\phi_- = \frac{\phi'_1 - \phi'_2}{\sqrt{2}}.
\label{eq:rot2}
\end{equation}
The densities associated with these new fields are defined as
\begin{equation}
\rho_{\pm} = \frac{1}{2\pi} \partial_x \phi_{\pm}
\label{eq:fipm}
\end{equation}
and they are related to the densities of the $\varphi'_{1}$ and $\varphi'_{2}$
fields by
\begin{equation}
\left(\matrix{\rho'_{1}\cr \rho'_{2}}\right)
=
\sqrt{\frac{g'}{2}} \left(\matrix{
1    & 1 \cr
1    & -1}\right)
\left(\matrix{
\rho_+\cr \rho_-} \right)
\label{eq:rot3}
\end{equation}
The successive rotations are schematized in Fig.(\ref{fig:fig-rotations}).

In terms of the $\phi_{\pm}$ fields the total Lagrangian reads:
\begin{eqnarray}
{\cal L} &=& \frac{1}{4\pi} \partial_x \phi_+ (\partial_t - \partial_x) \phi_+
+ \frac{1}{4\pi} \partial_x \phi_- (\partial_t - \partial_x) \phi_-  \nonumber
\\
         &+& \Gamma \delta(x) e^{-i \omega_0 t} e^{\sqrt{\frac{2}{g'}}
\phi_-(x,t)}+h.c. \label{eq:lagpm}
\end{eqnarray}
The strong coupling limit of this system is considerably simpler in the
dual picture described by the dual fields ${\tilde \phi}_{\pm}$. In terms of
the
dual fields the effective Lagrangian has a new Luttinger parameter
${\tilde g}'=1/g'$ and an effective tunneling amplitude ${\tilde \Gamma}\sim
\Gamma^{-{\frac{1+\nu}{2\nu}}}$. The (dual) Lagrangian is
\begin{eqnarray}
{\tilde {\cal L}} &=& \frac{1}{4\pi} \partial_x {\tilde \phi}_+ 
(\partial_t - \partial_x) {\tilde \phi}_+ 
+ \frac{1}{4\pi} \partial_x {\tilde \phi}_-
(\partial_t - \partial_x) {\tilde \phi}_-  \nonumber \\
                  &+& {\tilde \Gamma} \delta(x) e^{-i \omega_0 t}
e^{\sqrt{2g'} {\tilde \phi}_-(x,t)}+h.c.
\label{eq:lagpm1}
\end{eqnarray}

The dual transformation in the strong coupling limit ($\Gamma
\rightarrow \infty$ or ${\tilde \Gamma} = 0$) can be expressed in
terms of the fields $\phi'_1$ and $\phi'_2$:
\begin{eqnarray}
\phi'_1 &=&
\tilde\phi'_1 \Theta(-x) + \tilde\phi'_2 \Theta(x) \nonumber\\
\phi'_2 &=&
\tilde\phi'_2 \Theta(-x) + \tilde\phi'_1 \Theta(x)\ .
\label{eq:dualfi}
\end{eqnarray}
(here $\Theta(x)$ is the step function). The expectation values that
appear in the correlation functions are taken with respect to the dual
Lagrangian (with effective Luttinger parameter 1/g' and coupling constant
${\tilde \Gamma}$). As mentioned above, the duality transformation
relates the densities of the original fields and the densities of the
dual fields.  Notice that the densities in the incoming channels of
the original fields $\rho'_{1,2}(x<0)$, are the same as the densities
in the incoming channels of the dual fields ${\tilde
\rho}_{1,2}(x<0)$, {\it i.e.} the matrices given by Eqs. (\ref{eq:rot1},
\ref{eq:rot2}) can be used to express the original fields $\phi_{1,2}(x<0)$
in terms of the fields ${\tilde \phi}_{+,-}(x<0)$.  In order to write
the densities in the outgoing channels of the original fields
$\rho_{1,2}(x>0)$ in terms of the fields ${\tilde \phi}_{+,-}(x>0)$ it
is necessary to realize that the duality transformation exchanges
$\phi'_1$ and $\phi'_2$ for $x > 0$. As a consequence, for $x>0$
Eq.(\ref{eq:rot1}) reads:
\begin{equation}
\left(\matrix{\rho_1 \cr \rho_2}\right)
=
\left(\matrix{
\frac{1 + \sqrt{\nu}}{2}    & \frac{\sqrt{\nu} - 1}{2}\cr
\frac{1 - \sqrt{\nu}}{2\sqrt{\nu}} & \frac{1 + \sqrt{\nu}}{2\sqrt{\nu}}}\right)
\left(\matrix{
\tilde\rho'_{2}\cr \tilde\rho'_{1} } \right)
\label{eq:rot11}
\end{equation}

\begin{figure}
\vspace{.2cm}
\noindent
\hspace{.5in}
\epsfxsize=2.5in
\epsfbox{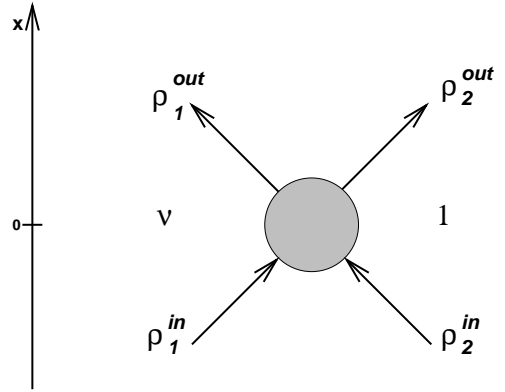}
\vspace{0.1cm}
\caption{Four probe geometry for measurement of noise.
The incident currents are those at $x < 0$ while the
outgoing currents are at $x > 0$.}
\label{fig:fig-scattering}
\end{figure}

Let us now use this formulation of the junction to calculate the
current correlation functions, necessary to compute both the current
through the junction and the noise. As shown in
Fig.~(\ref{fig:fig-scattering}) there are incoming ($x<0$) and
outgoing ($x>0$) scattering states with respect to the impurity
location ($x=0$).
The current flowing from the reservoir (or $\nu=1$ state) to the
filling fraction $\nu$ FQH state, can be written in terms of the
imbalance of the densities before ($x=0_{-}$) and after the impurity
($x=0_{+})$:
\begin{equation}
I=\rho_1^{\rm out}-\rho_1^{\rm in}=\rho_2^{\rm in}-\rho_2^{\rm out}\ \ 
\label{eq:currentofrho}
\end{equation}
Likewise, the current $I'$ and the densities $\rho'$ of the rotated fields
are related by
\begin{equation}
I'=\rho'^{\rm out}_1-\rho'^{\rm in}_1=\rho'^{\rm in}_2-\rho'^{\rm out}_2\ \ 
\label{eq:currentofrho'}
\end{equation}
As a consequence of current conservation and crossing symmetry (see
Fig.~\ref{fig:fig-scattering}) it follows that
\begin{equation}
I=I'
\label{eq:current-identity}
\end{equation}
Notice that Eqs.\ (\ref{eq:currentofrho}-\ref{eq:current-identity}) are 
operator identities and  not just relations between quantum averages.
 
The quantum noise for the current $I$, $S_I(\omega)$, is defined to be
\begin{equation}
S_I(\omega)\equiv  \int_{-\infty}^{\infty} dt\;\cos\omega t\; 
\langle \{I(t) , I(0)\} \rangle\ \ 
\label{eq:noiseofcurrent}
\end{equation}

The noise $S_I(\omega)$ defined in Eq.~(\ref{eq:noiseofcurrent}) is, in
general, a function of the tunneling amplitude $\Gamma$, the voltage
$V$ and the temperature $T$.  For the purpose of clarifying the
physics, we will separate the contributions to the noise into an {\it
equilibrium} piece, $S_I^{\rm eq}$, and an {\it excess} piece, $S_I^{\rm
ex}$. The latter, as the name suggests, is the amount by which the
noise, in the presence of a voltage $V$, exceeds the equilibrium
($V=0$) level.

\begin{figure}
\vspace{.15cm}
\noindent
\hspace{.5in}
\epsfxsize=2.5in
\epsfbox{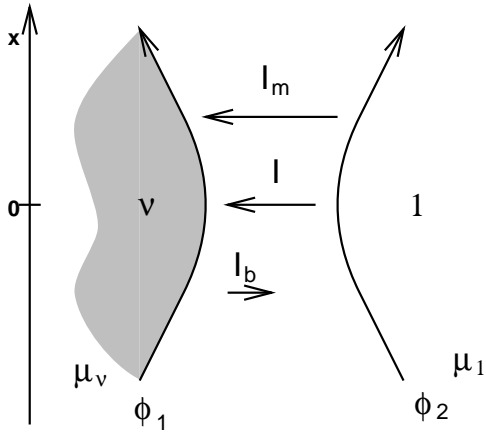}
\vspace{0.5cm}
\caption{Backscattering current $I_b$. $I$ is the current
through the junction for $\Gamma \neq 0$ and $I_m$ is the maximum
current through the junction.}
\label{fig:fig-current}
\end{figure}

Along the same lines, we can study the noise in the backscattering
current. The backscattering current $I_b $ is defined by (see
Fig.~\ref{fig:fig-current}):
\begin{equation}
I_b = I_m - I  
\label{eq:ib}
\end{equation}
where the current $I_m$ is the limiting value of $I$ in the strong
coupling (or large voltage) regime,{\it i.\ e.\/}, the maximum current
through the junction. Because of the operator identity between $I$ and
$I'$ (the current in the rotated system), $I_m=I'_m$. $I_m$ has the
following properties
\cite{K&F-noise,fls,FS2,LS3,CFW1,CFW,KF-dual,cl+ed}:
its mean value is $\langle I_m \rangle = g'\; {\frac{e^2}{h}}\;V$ and it
is a dissipationless current ,{\it i.\ e.\/}, its noise spectrum is completely
determined by the value of the conductance:
\begin{equation}
S_{I_m}(\omega) = g'{\frac{e^2}{2\pi}} |\omega|
\coth{\frac{|\omega|}{2T}}\ \ .
\label{eq:sim}
\end{equation}

From the definition of $I_b$ and the identity $I_m=I'_m$, it follows
that $I_b=I'_b$. Its noise is defined naturally as
\begin{equation}
S_{I_b}(\omega)\equiv \int_{-\infty}^{\infty} dt\;\cos\omega t\; 
\langle \{I_b(t) , I_b(0)\} \rangle\ \ 
\label{eq:noiseofcurrentIb}
\end{equation}

%


\section{Shot noise for a mismatched junction}
\label{sec:mismatch}

The procedure outlined in the previous section can be generalized for
a junction between two FQH liquids with filling fractions $\nu_1,
\nu_2$. In this case the {\it effective} filling fraction $g'$ takes the
value:
\begin{equation}
g'^{-1} = \frac {1}{2} (\nu_1^{-1} + \nu_2^{-1})
\label{eq:gralg'}
\end{equation}
{\it i.e.}, it is the harmonic average of both filling fractions.  

Recall that the tunneling current is unaffected by the rotation,  
\begin{equation}
I=I'\ \ 
\end{equation}
Since this is an operator identity, it follows that the noise in
this current is also the same in both the original and the rotated
problem (with the {\it effective} Luttinger parameter $g'$), hence
$S_I=S'_{I'}$. 

Let us focus now on the tunneling current $I$ and analyze its noise
$S_I$ in the two extreme regimes of weak and strong coupling, by
directly applying the known results for Luttinger liquids with the
same parameter \cite{K&F-noise,fls,FS2,LS3,CFW1,CFW}.

For the rest of this section we will only consider the shot noise,
namely the static $\omega=0$, zero temperature behavior of the noise.

\subsection{Shot noise in the weak coupling or strong backscattering
regime}

This is the limit where we can apply our physical intuition easily.
Here, the tunneling current between the distinct FQH liquids is
carried by electrons, the only common carriers between the decoupled
Laughlin states. This can be checked promptly by also looking at the
effective or rotated problem of tunneling between two Luttinger
liquids with $g'$ as given by Eq.~(\ref{eq:gralg'}). In the weak
tunneling limit of the problem, the relation between the zero
frequency shot noise level $S'_{I'}$ and the current $I'$ is $S'
=2 e\;I'$. Using the correspondence between the original problem and
the rotated one, we recover the intuitive result
\begin{equation}
S=2 \;e\;I
\end{equation}

Once we have checked this simple case, let us consider next the
non-trivial problem of strong coupling.

\subsection{Shot noise in the strong coupling or weak backscattering
regime}

It is in this case that the use of well established results for
tunneling between chiral Luttinger liquids with the same parameter
$g'$ is fundamental. The strong coupling limit of the problem
correspond to a dual system which can be treated in the weak coupling
limit. The relation of  noise and  current in this
limit is $S'_{I'}=2\;(g' e)\;I'_b$, where $I'_b=I'_m-I'$ is the
backscattering current, or deviations from the large voltage (or large
coupling) asymptotic current $I'_m=I_m=g'\;\frac{e^2}{h}\;V$. Such expression
implies that the noise-current relation for the mismatched FQH junctions is
\begin{equation}
S=2\;(g'e)\;I_b \ \ 
\end{equation}

This result can be interpreted as a consequence of uncorrelated or
Poissonian tunneling events of fractionally charged carriers of charge
$g'$ given by the harmonic average of the filling factors
$\nu_1,\nu_2$ on the two sides of the junction. The important question
is which carriers have such charge. The state of charge $g' e$ does
not exist in either isolated FQH system, be it in the bulk or in the
edge. Such state is a soliton of the {\it strongly coupled} edges of
the two FQH states. The shot noise suppression factor should be a
measure of the charge of these soliton states, which in general differ
from the charges of the Laughlin quasiparticles. 

An important special case is that of $\nu_1=\nu_2$. Here, the charge
of the soliton state for the strongly coupled FQH edges is the same as
that of the quasiparticle states or solitons constructed in the
isolated system. In general, the soliton quantum numbers are a
property of the coupled system as a whole.

\section{Exact solution for tunneling between a Fermi Liquid and
a $\nu=1/3$ FQH state}
\label{sec:noiset0}

In this section we present the results for the noise $S_I$ and $S_{I_b}$
in a junction between a normal metal and a $1/3$ FQH state at zero
temperature. This case corresponds to an exactly solvable point which
can be studied via refermionization for the entire range of couplings,
voltages and temperatures. In this section we discuss the $T=0$ case
and devote the next section for finite temperature effects.

\subsection{Refermionization}
\label{sec:free}

For a junction between a normal metal and a FQH system with filling
fraction $\nu = 1/3$, the value of the {\it 'effective'} filling
fraction is $g' = 1/2$.  Thus, for $g'=1/2$ the coefficient in front
of ${\tilde \phi}_-$ in the tunneling term is $\sqrt{2g'} = 1$. In
this particular case the effective Hamiltonian can be diagonalized by
defining a fermion operator $\psi(x,t) \propto : e^{i {\tilde
\phi}_-(x,t)} :$. As shown in Ref.~\cite{CFW}, the diagonalization
carried out through this refermionization procedure allows an exact
solution for {\it all} values of ${\tilde
\Gamma}$ . Thus, for a junction $(1,{\frac{1}{3}})$ a full
solution for the correlation functions and hence the noise spectrum
can be obtained.

The fermionic fields that diagonalize exactly the Hamiltonian are
given by \cite{CFW}:
\begin{equation}
\psi(x) = \cases{\sum_\omega A_\omega e^{i(\omega + \omega_{qp})x}
e^{-i\omega t}                       & for $x < 0$
\label{psixminus}\cr
                 \sum_\omega B_\omega e^{i(\omega + \omega_{qp})x}
e^{-i\omega t}                       & for $x > 0$
\label{eq:psixplus}\cr}
\end{equation}
and
\begin{equation}
\psi^\dagger(x) = \cases
     {\sum_\omega A^\dagger_{-\omega} e^{i(\omega - \omega_{qp})x}
e^{-i\omega t}                       & for $x < 0$
\label{psidxminus}\cr
     \sum_\omega B^\dagger_{-\omega} e^{i(\omega - \omega_{qp})x} e^{-
i\omega t}                       & for $x > 0$,
\label{eq:psidxplus}\cr}
\end{equation}
where
\begin{equation}
B_\omega = {(1 + e^{i\phi(\omega)})A_\omega
           + (1-e^{i\phi(\omega)})A^\dagger_{-\omega} \over 2},
\label{eq:Bdef}
\end{equation}
and
\begin{equation}
e^{i\phi(\omega)} = {i\omega + 4\pi|{\tilde \Gamma}|^2 \over i\omega -
4\pi|{\tilde \Gamma}|^2 }.
\label{eq:phidef}
\end{equation}
The commutation relations obeyed by the operators $A_{\omega}$ are:
\begin{equation}
\{A_{\omega_1}, A^\dagger_{\omega_2}\} = \delta_{\omega_1,
\omega_2}.
\label{eq:Acom}
\end{equation}
The scattering state $| \Phi \rangle$ incident upon the junction is in
equilibrium with the reservoir (the normal metal side of the
junction), which has energy $\omega_{qp} = (1/2)(eV/\hbar)$. At zero
temperature all the states with energies $\omega < \omega_0$ are
filled. This implies:
\begin{eqnarray}
A^\dagger_\omega |\Phi \rangle &=& 0 \qquad {\rm for} \quad \omega <
\omega_{qp}  \nonumber \\
A_\omega |\Phi \rangle         &=& 0 \qquad {\rm for} \quad \omega >
\omega_{qp}.
\label{eq:aphi}
\end{eqnarray}
As a consequence of the anticommutation relations from
Eq.~(\ref{eq:Acom}) we have,
\begin{eqnarray}
 \langle \Phi | A_{\omega_1} A_{\omega_2} |\Phi \rangle       &=& 0 \nonumber\\
\langle \Phi | A^\dagger_{\omega_1} A_{\omega_2}|\Phi \rangle &=&
n_{\omega_1}\delta_{\omega_1, \omega_2},
\label{eq:AdAexp}
\end{eqnarray}
where at temperature $T = 0$
\begin{equation}
   n_\omega = \cases{1 & for $\omega < \omega_{qp}$ \cr
                     0 & for $\omega > \omega_{qp}$. \cr}
\label{eq:nzT}
\end{equation}
and for $T \neq 0$ it corresponds to the Fermi distribution function with
$\omega_{qp}$ as the chemical potential.
The density ${\tilde \rho}_- = (1/2\pi) \partial_x{\tilde \phi}_- $
can be written in terms of these fermionic fields as ${\tilde \rho}_-(x) =
\psi^{\dagger}(x) \psi(x)$, so that all correlation functions of
${\tilde \rho}_-$ can be derived from the correlations of the
fermions.

\subsection{Noise in current $I$}

The current correlations can be calculated by expressing the current
operator in terms of the densities $\tilde\rho_\pm$, the natural
quantities that appear in the exact solution. The current $I$ through
the junction is given by Eq.~(\ref{eq:currentofrho}), which we repeat
below for the sake of completeness:
\begin{equation}
I=\rho_1^{\rm out}-\rho_1^{\rm in}=\rho_2^{\rm in}-\rho_2^{\rm out}\ \ 
\end{equation}

By using Eqs.(\ref{eq:rot1}), (\ref{eq:rot3}) and (\ref{eq:rot11}) the
densities $\rho_{1,2}$ can be written in terms of the densities of the
fields ${\tilde \phi}_{+,-}$:
\begin{eqnarray}
\rho_1(x=0_-, t) &=& \sqrt{\frac{g'\nu}{2}} {\tilde \rho}_+(0_-, t) +
\sqrt{\frac{g'}{2}} {\tilde \rho}_-(0_-, t)    \nonumber \\
\rho_2(x=0_-, t) &=& \sqrt{\frac{g'}{2\nu}} {\tilde \rho}_+(0_-, t) -
\sqrt{\frac{g'}{2}} {\tilde \rho}_-(0_-, t) \nonumber \\
&&
\label{eq:incnoi}
\end{eqnarray}
and
\begin{eqnarray}
\rho_1(x=0_+, t) &=& \sqrt{\frac{g'\nu}{2}} {\tilde \rho}_+(0_+, t) -
\sqrt{\frac{g'}{2}} {\tilde \rho}_-(0_+, t)    \nonumber \\
\rho_2(x=0_+, t) &=& \sqrt{\frac{g'}{2\nu}} {\tilde \rho}_+(0_+, t) +
\sqrt{\frac{g'}{2}} {\tilde \rho}_-(0_+, t) \nonumber \\
&&
\label{eq:outnoi}
\end{eqnarray}
The current is therefore given by
\begin{equation}
I=\sqrt{\frac{g'}{2}} \left[
{\tilde \rho}_-(0_+, t) + {\tilde \rho}_-(0_-, t)
\right]
\label{eq:Ifrom-}
\end{equation}

Notice that the contribution to $I$ due to $\tilde\rho_+$ drops out,
since this is a free field, and thus continuous across the
impurity. 

The backscattering current $I_b$ can be shown to be given by 
\begin{equation}
I_b=\sqrt{\frac{g'}{2}} \left[-{\tilde \rho}_-(0_+, t) + {\tilde \rho}_-(0_-, t)\right]
\label{eq:Ibfrom-}
\end{equation}
by using the definition of $I_b$ in terms of $I$ and $I_m$ given in
Eq.~(\ref{eq:ib}), and by satisfying the dissipationless property of
$I_m$ as expressed in Eq.~(\ref{eq:sim}). Alternatively, one can simply
use the fact we showed previously, that $I_b=I'_b$. Thus,
Eq.~(\ref{eq:Ibfrom-}) is the natural expression for $I_b$ in terms of
$\tilde\rho_-$.

It is convenient to define here $T_k$, the crossover energy scale set
by the tunneling amplitude, as $T_k = 4 \pi |{\tilde \Gamma}|^2$.  In
what follows, we will express the noise $S_I$ and $S_{I_b}$ in terms of $T_k$,
the frequency $\omega$ and the voltage $V$, through the Josephson
frequency $\omega_0=eV$ (in units with $\hbar=1$).  By dimensional
analysis, we expect the noise to be expressible, up to a scale factor,
in terms of a dimensionless function of the ratios
${\frac{\omega}{T_k}}$ and ${\frac{V}{T_k}}$.

We can now use Eqs.~(\ref{eq:Ifrom-}),(\ref{eq:Ibfrom-}) and the correlation
functions for ${\tilde \rho}_-$ calculated using the refermionized
version of the problem to obtain the noise $S_I(\omega)$. Because of the
definition of $I_b$ it is easy to check that $S_I(\omega) = S_{I_b}(\omega)$
at $T=0$.

\breakon

After some algebra we find,
\begin{equation}
S_I \left(\omega,V,T_k \right)= \frac{e^2}{2\pi }|\omega| +  g'\;
{\frac{e^2 }{2\pi}}  T_k \;\; \Theta(|eV| - |\omega|) \; 
\left[ \tan^{-1}
{\frac{|eV|}{2T_k}}  
+ \tan^{-1} {\frac{|eV| -2 |\omega|}{2T_k}}    
+ \frac{T_k^2}{|\omega|} 
\ln 
{\frac{
T_k^2 + (|\omega| - |eV|/2|)^2}
{T_k^2 +(eV/2)^2}}
\right] \ \ .
\label{eq:long1} 
\end{equation}

The DC shot noise takes the form:
\begin{equation}
S_I(0,V, T_k ) = 2 g'\;{\frac{e^2}{2\pi}} T_k
 \;\left[  \tan^{-1} 
{\frac{|eV|}{2T_k}}  \nonumber 
- 
{\frac{
{\frac{|eV|}{2T_k}} }{1+\left(
{\frac{eV}{2T_k}}\right)^2}}
 \right] 
\label{eq:long2}
\end{equation}
\breakoff 

Next we relate the expression for the noise with the expression for
the backscattering current $I_b$. After some algebra, the expression for
$\langle I_b \rangle$ is found to be given by:
\begin{equation}
\langle I_b \rangle = 2 T_k \sqrt{\frac{g'}{2}} {\frac{e}{2\pi}} \tan^{-1}
\frac{eV}{2T_k}
\label{eq:ibb}
\end{equation}

Notice that $\langle I_b \rangle$ goes to $0$ when $T_k \rightarrow 0$
($\Gamma \rightarrow \infty$: all the current incident upon the
junction is transmitted and nothing is reflected back), while it goes
to $I_m$ when $T_k \rightarrow \infty$ ($\Gamma \rightarrow 0$: all
the current incident upon the junction is reflected back) as
expected. Thus, the strong coupling regime $\Gamma \rightarrow \infty$
can be viewed alternatively as the weak backscattering regime .
Conversely, the weak coupling regime $\Gamma \rightarrow 0$ is also
the strong backscattering regime.

Finally the expression for the noise in terms of the current is given
by:
\begin{equation}
S_I( \omega=0,V,T_k ) = 2 e
{\sqrt{\frac{g'}{2}}}
\left(
 V \frac{\partial \langle I_b \rangle}{\partial V}-\langle I_b \rangle 
\right)
\label{eq:fsl}
\end{equation}

This is in complete agreement with results obtained in Ref.\cite{fls}.
As discussed in that work, this expression can be studied in the  
strong (${\tilde \Gamma}\rightarrow 0$) and weak coupling (${\tilde
\Gamma}\rightarrow \infty$) limits which, as mentioned above, correspond
respectively to the weak and strong backscattering regimes. Thus
\begin{eqnarray}
&&S_I(\omega=0, {\tilde \Gamma}\rightarrow 0) = 
2 \sqrt{\frac{g'}{2}} e \langle I_b \rangle =  2 \frac{e}{2} \langle I_b
\rangle
\nonumber \\
&&S_I(\omega=0, {\tilde \Gamma} \rightarrow \infty) = 2 e (I_m - \langle
I_b \rangle) =  2 e \langle I \rangle
\label{eq:wands}
\end{eqnarray}

The expression for the weak backscattering limit is the main result of
this work: the charge appearing in front of the backscattering current
is an {\it effective} charge given by $g'$. Since $g'=1/2$, the value
of this {\it effective} charge is $e/2$. However the original junction
is between a FQH system in the state $\nu = 1/3$ and a normal metal
(equivalent to a FQH system in the state $\nu = 1$) and hence there
are no quasiparticles with fractional charge $e^* =
{\frac{e}{2}}$. This implies that the scale of the shot noise spectrum
is not determined by the charge of the carriers present in the
decoupled system but by the charge of {\it effective} carriers. It is
also worth to mention that in this junction the value of the
conductance is given by $G = {\frac{1}{2}} {\frac{e^2}{\hbar}}$, {\it
i.\ e.\/} the conductance is also determined by the {\it effective}
filling fraction.

In addition to setting the scale of the zero-frequency shot noise, the
effective charge $e/2$ is also manifest in the finite frequency
spectrum. Two features emerge from Eq.(\ref{eq:long1}):
\begin{enumerate}
\item
The noise in the tunneling current vanishes beyond a frequency
$\omega_{\rm el}=eV$, set by the non-equilibrium voltage. This can be
readily read from the step function in Eq.~\ref{eq:long1}), and so
this is valid for any $T_k$. We regard this singularity as evidence
that the physical particles that tunnel are electrons
\cite{CFW}. Therefore, regardless of whether the tunneling is in the
weak or strong regime, this singularity in the noise spectrum will be
present at the ``electron'' frequency $\omega_{\rm el}=eV$.  This
result, the vanishing of the spectrum beyond the electron frequency,
is an exact result for this case $g'=1/2$, as well as for
non-interacting electrons, $g'=1$. However, it is unclear whether this
result should hold more generally (see Ref.~\cite{LS3}).

\item
In the limit of small $T_k \ll eV$, one finds from Eq.~\ref{eq:long1})
that there is structure (a smeared singularity) in the noise spectrum
at a frequency $\omega_{\rm qp}=\frac{e}{2}V$, corresponding to the
charge $e/2$ quasiparticles which are backscattered in the strong
coupling or weak backscattering regime. The frequency range over which
the singularity is smeared is set by the energy scale $T_k$ and it is
centered at $\omega_{\rm qp}=\frac{e}{2}V/h$. The presence of this
singularity in the spectrum  provides further evidence for the
existence of charge ${\frac{e}{2}}$ states.
\end{enumerate}

\section{Noise and current for a ($1,\frac{1}{3}$) junction at $T \neq 0$}
\label{sec:noiset}

In this section we calculate the DC noise $S_I$ in the current through
the junction $I$ and the noise $S_{I_b}$ in the backscattering current
$I_b$ at temperatures $T \neq 0$. We present the exact expression for
$S_I$, $S_{I_b}$ and $I_b$ for all values of voltage $V$,
temperature $T$ and energy scale $T_k$. We discuss, in particular,
asymptotic limits for both noise and current and summarize the results
in terms of a temperature-voltage ($T-V$) diagram. We also show the
existence of universal ratios between the two limit regimes of {\it
shot} and {\it thermal} noise that should be accessible experimentally.

The calculation for the noise and the current follows the same steps
as in the previous section, the only difference being the expression
used for the Fermi distribution function. For simplicity, in what
follows we will work in units of $\hbar=k_B = 1$.

\breakon

The general expressions for the noise $S_I$ in the current $I$ and the
noise $S_{I_b}$ in the backscattering current $I_b$, at finite
voltage, temperature and zero frequency are given by:
\begin{eqnarray}
S_I( V, T,T_k)&=&  \frac{e^2}{2\pi } g' T_k \left[4 \frac{T}{T_k} + F_1\left({\frac{V}{T_k}},{\frac{T}{T_k}}\right) -  F_2\left({\frac{V}{T_k}},{\frac{T}{T_k}}\right) \right]
\label{eq:silong} \\
S_{I_b}( V, T,T_k) &=& \frac{e^2}{2\pi } g' T_k \left[4 \frac{T}{T_k} + F_1\left({\frac{V}{T_k}},{\frac{T}{T_k}}\right) +  F_2\left({\frac{V}{T_k}},{\frac{T}{T_k}}\right) \right]
\label{eq:siblong}
\end{eqnarray}
where
\begin{eqnarray}
F_1 &=& \sinh\left(\frac{eV}{2T}\right) \int_{\-infty}^{\infty}\;\; dx \;\;\;\frac{x^2}{(1 + x^2)^2} \;\;\frac{\tanh\left(\frac{eV}{4T} - \frac{xT_k}{2T}\right)}
{\cosh\left(\frac{eV}{4T} + \frac{xT_k}{2T}\right) \cosh\left(\frac{eV}{4T} - \frac{xT_k}{2T}\right)} 
\label{eq:f1} \\
F_2 &=& \int_{-\infty}^{\infty}\;\; dx \;\;\;\frac{1}{1 + x^2} \;\;\frac{1}{\cosh^2\left(\frac{eV}{4T} - \frac{xT_k}{2T}\right)}
\label{eq:f2} 
\end{eqnarray}
\breakoff

The exact expression for the functions $F_{1,2}$ are calculated in
appendix \ref{sec:app1}.

The first term in the expressions for $S_I$ and $S_{I_b}$ is the
expected equilibrium noise result, corresponding to the
Johnson-Nyquist (thermal) noise. The
dimensionless functions
$F_{1,2}\left({\frac{V}{T_k}},{\frac{T}{T_k}}\right)$ are scaling
functions which describe the crossover between the weak tunneling
fixed point at $\Gamma \to 0$, and the strong tunneling fixed point at
$\Gamma \to \infty$. For this integrable model of the junction, these
scaling functions are universal and depend only on one parameter, the
crossover scale $T_k$. The crossover between the two asymptotic
regimes is then controlled by the voltage $V$ and the temperature $T$.
The weak tunneling regime is accessed in the limit of low temperature
($T \ll T_k$) and low voltage ($V \ll T_k$). Conversely, the strong
coupling regime can be accessed either at high temperatures and low
voltages, or at large voltages and low temperatures. The {\it shot
noise} regime corresponds to temperatures lower than the applied external
voltage, {\it i.e.}, $T \ll V$ and the {\it thermal noise} regime
corresponds to applied voltages lower than the temperature,{\it i.e.}, $V
\ll T$. These last two regimes are interpolated smoothly as a function
of ${\frac{V}{T}}$.

Finally, it is also possible to obtain an exact expression for the
backscattering current at finite temperatures and voltages. Repeating
the procedure outlined in the previous section for the calculation of
$\langle I_b \rangle$ we obtain:
\begin{equation}
\langle I_b \rangle = \frac{1}{2} \frac{e}{2\pi} \sqrt{\frac{g'}{2}} \, T_k \,\, \sinh\left(\frac{V}{2T}\right)
F_3\left({\frac{eV}{T_k}}, {\frac{T}{T_k}}\right)
\label{eq:scalingi}
\end{equation}
where
\begin{equation}
F_3 = \int_{-\infty}^{\infty} {\frac{dx}{(1 + x^2)}}
{
\frac{1}
{\cosh\left(\frac{eV}{4T} -\frac{xT_k}{2T}\right)
\cosh\left(\frac{eV}{4T} + \frac{xT_k}{2T}\right)}
}
\label{eq:F3}
\end{equation}

\subsection{Asymptotic limits and universal noise ratios}

\begin{figure}
\vspace{.2cm}
\noindent
\hspace{.5in}
\epsfxsize=2.5in
\epsfbox{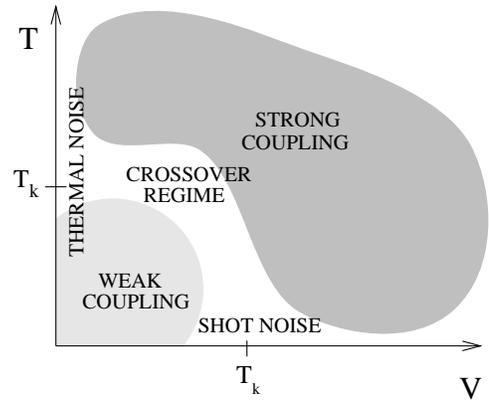}
\vspace{-2.0cm}
\caption{Phase diagram for $S_I$ and $S_{I_b}$. The strong coupling regime
is characterized by either high temperature or high voltage
values. The weak coupling regime corresponds to low temperature, low
voltage values. $T_k$ is the energy scale that determines the
crossover regime. Thermal and shot noise are characterized by the 
regions $T \gg V$ and $V \gg T$ respectively.}
\label{fig:pdn}
\end{figure}

A better understanding of the roles played by temperature and voltage
results from analyzing the T-V diagram for both $S_I$ and $S_{I_b}$,
focusing on the behavior of the function $F_{1,2}$ in different
limiting regimes as shown in Fig.\ref{fig:pdn}.

There are two interesting regimes: Thermal noise or $T \gg V$ and
Shot noise or $V \gg T$.

\begin{itemize}
\item
$T \gg V$ Thermal noise

\begin{equation}
S_I =  4 g' \frac{e^2}{2\pi }  T - S_{I_b} 
= 
\cases{
g' \frac{e^2}{2\pi }\frac{4}{3} {\pi}^2 T_k \left(\frac{T}{T_k}\right)^3  &\mbox{$T_k \gg T $}
\cr       
\cr
4 g'  \frac{e^2}{2\pi } T - 2 g' \frac{e^2}{2\pi } \pi T_k &\mbox{$T_k \ll T $}\cr }
\label{eq:thermal}
\end{equation}

\item
$V \gg T$ Shot noise

\begin{equation}
S_I = S_{I_b} 
= 
\cases{
g' \frac{e^2}{2\pi } \frac{4}{3} T_k \left(\frac{eV}{2T_k}\right)^3  &\mbox{$T_k \gg  V $}
\cr       
\cr
2 g' \frac{e^2}{2\pi } \pi T_k   &\mbox{$T_k \ll V $}\cr 
 }
\label{eq:shot}
\end{equation}
\end{itemize}
(At $V=0$ the function $F_1=0$ and the limiting behavior of the
function $F_2$ is calculated in appendix \ref{sec:app2}.)

Furthermore, for $T_k \ll T$, $I_b \ll I$ (weak backscattering
regime), and the expression for the noise in the current $I$ takes the form
\begin{equation}
S_I = 4 g' \frac{e^2}{2\pi } T + 2 g' \frac{e^2}{4\pi } \pi T_k \left[1 -\frac{2}{\cosh^2\left(\frac{eV}{4T}\right)}\right]
\end{equation}

From this expression, the crossover between thermal and shot noise can be
studied. The crossover region is determined by the argument of
the $\cosh(eV/4T)$ function. If the argument is written in the standard form
of $(QV/2T)$, the value of $Q$ determined by this crossover corresponds to 
$Q = e/2$.

The backscattering current $I_b$ can also be calculated in this regime. 
Its expression reads:
\begin{equation}
I_b=\frac{e}{2\pi } \sqrt{\frac{g'}{2}} T_k \pi \tanh\left(\frac{eV}{4T}\right)
\end{equation}

After some algebra, the noise $S_I$ can be put in terms of the current $I_b$
\begin{equation}
S_I = 4\,G\,T +  2 \left(\frac{e}{2}\right) I_b \tanh^{-1}\left(\frac{eV}{4T}\right) - 4 T \frac{dI_b}{dV}
\end{equation}
where $G = dI/dV$ is the differential conductance of the
junction. This expression has been obtained in previous works by Kane and
Fisher~\cite{K&F-noise} and Fendley and Saleur~\cite{FS2}.

Further information can be obtained by comparing the limiting behavior
of the noise in both, thermal and shot noise regime.  In the strong
coupling regime $T \gg T_k $ or $V \gg T_k $ the noise $S_{I_b}$ {\it
saturates} to the same constant value for both cases, equilibrium ($
V=0$) and non-equilibrium ($V \neq 0$). Notice that, while the
saturation values of the noise $S_{I_b}$ both in the thermal and shot
noise limits is determined by the (non-universal) energy scale $T_k$,
their ratio takes the universal value
\begin{equation}
R_{I_b}={\frac{S_{I_b}(T=0;  T_k \ll T)}{S_{I_b}(V=0; T_k \ll V)}}=1
\label{eq:ratio1}
\end{equation}

Similarly, in the weak coupling regime $T_k \gg T $ or $T_k \gg V $,
the noise $S_I$ {\it vanishes} as $(\frac{eV}{T_k})^3$ in the
non-equilibrium regime ($V \neq 0$) and as $(\frac{T}{T_k})^3$ in the
equilibrium one ($V =0$), as expected since the tunneling operator is
irrelevant in the $\Gamma=0$ fixed point. Thus, the weak coupling
regime can also be characterized by a ratio that is independent of the
energy scale $T_k$. In this case, the ratio involves the noise $S_I$
in both the thermal and shot noise limits
\begin{equation}
R_I=
{\frac{S_I (T=0;  T_k \gg T)}{S_I (V=0; T_k \gg V)}}= \frac{1}{8 {\pi^2}} \left(\frac{eV}{T}\right)^3
\label{eq:ratio2}
\end{equation}

It is worth to stress that while in the expression for $S_I$ and $S_{I_b}$,
temperature and voltage play an analogous role, their physical meaning
is quite different since in the $V=0, T\neq 0$ the system is in
thermal equilibrium, while for $V\neq 0, T=0$ it is away from
equilibrium.

\section{Generalized Universal Noise Wilson Ratios}
\label{sec:ratios}

The ratios introduced above are interesting from the point of view of
theory and experiment. These quantities are {\it universal amplitude
ratios} of the behavior of the noise level at a given fixed point of
the junction. They are a universal property of the fixed point. As
such they are determined without a detailed knowledge of the crossover
region. Hence, they can also be determined by a direct study of the
behavior of the noise in the vicinity of the fixed point of interest,
even for non-integrable systems. In particular, a perturbative
analysis of the behavior of a generic $(\nu_1,\nu_2)$ junction can be
used to calculate these ratios for a more general case. This is the
purpose of this section.

Universal amplitude ratios are common in critical systems in general
and in quantum impurity problems in particular. In the case of Kondo
systems, the Wilson ratio, the ratio of the (suitably normalized)
impurity paramagnetic susceptibility and of the slope of the impurity
specific heat, describes the approach to the fixed point. The ratio
defined in the last section bears an obvious similarity with the
Wilson ratio.  

Experimentally, both the thermal and shot noise can be measured, and
therefore their ratio can be obtained and used to test the theoretical
prediction based on the Luttinger liquid model for the edge states.

The asymptotic values of both the thermal and shot noise levels can be
obtained, using perturbation theory, for a general junction
$(\nu_1,\nu_2)$ through the value of the effective Luttinger parameter
$g'$. In the perturbative approach the expansion is done around a
particular fixed point. For the weak coupling fixed point $\Gamma
\rightarrow 0$ the current $I$ through the junction is small compared
to the asymptotic large voltage current $I_m$. On the other hand, for
the strong coupling $\Gamma \rightarrow
\infty$, the backscattering current $I_b$ is small. Thus the ratios
can be calculated in both cases by focusing on the appropriate small
current. The duality relation between the two fixed points allows one to
calculate the noise in both currents by simply taking $g'$ into
$1/g'$.

\breakon

The perturbative results are given by (see Ref.~\cite{CFW1})  
\begin{eqnarray}
S_I    &=& 2 e^2 |\Gamma|^2 (2\pi T)^{\frac{2}{g'}-1} \;B 
\left(\frac{1}{g'} + i\frac{eV}{2\pi T},\frac{1}{g'}  
- i\frac{eV}{2\pi T}\right)
\cosh\left(\frac{eV}{2T}\right)
\label{eq:si1}  \\
S_{I_b}&=& 2 {e^*}^2 |\tilde\Gamma|^2 (2\pi T)^{2g'-1} \;B 
\left(g' + i\frac{e^*V}{2\pi T},g'- 
i\frac{e^*V}{2\pi T}\right)
\cosh\left(\frac{e^*V}{2T}\right)
\label{eq:sib1}
\end{eqnarray}
where $e^*=g' e$, the charge of the quasiparticle responsible for the
noise in the weak backscattering limit. 
\breakoff

The shot to thermal noise ratios are then given by:
\begin{eqnarray}
R_I &=& 
\frac{S_{I}(T\to 0)}{S_{I}(V\to 0)}=
\frac{\pi}{\Gamma^2(1/g')}\;\left(\frac{eV}{2\pi T}\right)^{2/g'-1} \\
R_{I_b} &=&
\frac{S_{I_b}(T\to 0)}{S_{I_b}(V\to 0)}=
\frac{\pi}{\Gamma^2(g')}\;\left(\frac{e^* V}{2\pi T}\right)^{2g'-1}
\end{eqnarray}

In particular, for $g'=1/2$, we obtain 
\[
R_I=\frac{1}{8\pi^2}
\left(\frac{eV}{T}\right)^3
\qquad {\rm and} \qquad R_{I_b}=1
\]
which are the results obtained non-perturbatively in the previous
section.

For a junction $(\frac{1}{3},\frac{1}{3})$ the effective Luttinger parameter $g' = 1/3$ and
these ratios are
\begin{eqnarray}
R_I     &=& \frac{S_{I}(T\to 0)}{S_{I}(V\to 0)} =\frac{1}{128 {\pi}^4} \left(\frac{eV}{T}\right)^5    \nonumber \\
R_{I_b} &=& \frac{S_{I_b}(T\to 0)}{S_{I_b}(V\to 0)}=\left[\frac{\pi (6 \pi )^{\frac{1}{3}}}{\Gamma^2(\frac{1}{3})}\right] \left(\frac{T}{eV}\right)^{\frac{1}{3}}
\end{eqnarray}

Thus, these expressions could be compared with noise measurements in
the geometries used by L.~Saminadayar et.~al \cite{Glattli} and by
R.~de Picciotto et.~al\cite{Reznikov} in the shot and thermal noise
regimes. 

These ratios are universal scaling functions in the sense that they
are independent of the coupling constant ($\Gamma$ or $\tilde\Gamma$)
or, alternatively, of the energy scale $T_k$. They still contain
information on the Luttinger liquid behavior of the edge states which
is manifest in the $V/T$ exponent and the $g'$ dependent prefactor.

\section{Auto-correlations in the incoming and outgoing branches}
\label{sec:auto-correlations}

Thus far we have focused on the noise in the current flowing through
the junction. More information can still be extracted from the noise
by looking at correlations in the four branches involved in the
scattering separately. In particular, the noise on both currents, $I$
and $I_b$, can be obtained from the measurements of correlations in a
four probe setting, as done by L.~Saminadayar {\it et al.}
\cite{Glattli}.

For this purpose, let us define the following auto and cross
correlations between the densities in the various branches:
\begin{equation}
S_{ij}(\omega;x,y) \equiv \int_{-\infty}^{\infty} dt
\;\cos\omega t\; 
\langle \{\rho_i(t,x) , \rho_j(0,y)\} \rangle.
\label{eq:sijofomega}
\end{equation}
where the $i,j$ subscripts labels the branches $1,2$ (see
Fig.~\ref{fig:fig-scattering}). In this section we will focus
primarily on the correlations for branches on the same side of the
impurity ($x=y$).  Consequently, we will work with
\begin{eqnarray}
S^{\rm in\ }_{ij}(\omega)&=& S_{ij}(\omega;x,y) \quad {\rm for}\quad
x=y<0
\label{eq:def-in}\\
S^{\rm out}_{ij}(\omega)&=& S_{ij}(\omega;x,y) \quad {\rm for}\quad  x=y>0
\label{eq:def-out}
\end{eqnarray}
according to the side where the densities in consideration lie with
respect to the impurity. Whenever we wish to address a general
property of the noise valid in either case, we shall also drop the
superscripts $out$ and $in$, and simply use $S_{ij}$.

Just as in the case of $S_I(\omega)$ defined in
Eq.~(\ref{eq:noiseofcurrent}), the related quantities defined in
Eqs.~(\ref{eq:sijofomega}-\ref{eq:def-out}) are also functions of the
tunneling amplitude $\Gamma$, the voltage $V$ and the temperature $T$.
Besides the separation between an {\it equilibrium} $S^{\rm eq}$ and
an {\it excess} $S^{\rm ex}$ contribution to the noise, we can
alternatively also separate the noise into an {\it impurity}
contribution $S^{\rm imp}$, and the contribution of the {\it two
decoupled channels}, $S^{\Gamma=0}$. This separation of physical
quantities is standard in quantum impurity problems. $S^{\rm imp}$ and
$S^{\Gamma=0}$ are simply related to quantities easily calculable in
the basis of the fields ${\tilde \phi}_{+,-}$. Since these fields are
decoupled from each other, we only need to calculate their
auto-correlation functions. The field ${\tilde \phi}_+$ is a free
field, and its noise spectrum at zero temperature can be calculated in
a straightforward way giving:
\begin{equation}
S_+(\omega) = {\frac{e^2}{2\pi}} |\omega| \coth{\frac{|\omega|}{2T}}
\label{eq:sp}
\end{equation}
The noise spectrum for the field ${\tilde \phi}_-$ is defined
similarly as above to be:
\begin{equation}
S_-(\omega; x, y) = \int_{-\infty}^{\infty} dt\; \cos\omega t\;
\langle
\{\rho_-(t, x) , \rho_-(0, y)\} \rangle.
\label{eq:s-omega}
\end{equation}

Below, we give explicit relations between $S_\pm(\omega)$ and the
noise expressions $S^{\rm imp}$ and $S^{\Gamma=0}$ for the incoming
and outgoing branches.

The noise in the original branches can be related to the noise in the
rotated branches, as follows.  We begin by utilizing the
generalization of Eq.~(\ref{eq:rot1}) for two filling fractions
$\nu_1$ and $\nu_2$:
\begin{equation}
\left(\matrix{\rho_1 \cr \rho_2}\right)
=
\frac{1}{2}\;
\left(\matrix{
\sqrt{\frac{\nu_1}{\nu_2}}+1    & \sqrt{\frac{\nu_1}{\nu_2}}-1 \cr
\sqrt{\frac{\nu_2}{\nu_1}}-1 & \sqrt{\frac{\nu_2}{\nu_1}}+1}
\right)
\left(\matrix{
\rho'_{1}\cr \rho'_{2} } \right)\ .
\label{eq:genrot1}
\end{equation}

In order to calculate the noise spectrum of this junction we use
Eq.~(\ref{eq:sijofomega}). The noise, or density, correlations in the
two problems can also be related through the matrix in
Eq. (\ref{eq:genrot1}). It suffices to look at $S_{11}$, since the
results are analogous for $S_{22}$.
\begin{equation}
S_{11}=
\frac{1}{2}
\left[
\left(\frac{\nu_1}{\nu_2}+1\right) \; S'_{11}
+
\left(\frac{\nu_1}{\nu_2}-1\right)
\; S'_{12}
\right]\ \ ,
\label{eq:SS'}
\end{equation}
where we used that $S'_{11}=S'_{22}$ and $S'_{12}=S'_{21}$.

Now we need a relationship between $S'_{12}$ and $S'_{11}$, which we
obtain by rotating to the decoupled $\phi_\pm$ fields as in
Eq. (\ref{eq:rot3}) (which we repeat below for completeness; see also
Fig. \ref{fig:fig-rotations}):
\begin{equation}
\left(\matrix{\rho'_{1}\cr \rho'_{2}}\right)
=
\sqrt{\frac{g'}{2}} \left(\matrix{
1    & 1 \cr
1    & -1}\right)
\left(\matrix{
\rho_+\cr \rho_-} \right)\ \ .
\end{equation}

We can write
\begin{eqnarray}
S'_{11} &=& \frac{g'}{2}
\left( S_{+}+S_{-} \right)\\
S'_{12} &=& \frac{g'}{2}
\left( S_{+}-S_{-} \right)
=\frac{g'}{2}\left( 2 S_{+}\right) -
\frac{g'}{2} \left( S_{+}+S_{-} \right) \nonumber \\
&=& S'^{\Gamma=0}_{11} - S'_{11} \ \ .
\end{eqnarray}

We have used in the last line above the fact that, at $\Gamma=0$,
 $S_{-}=S_{+}$. Thus, one can cast one of the terms
as $S'^{\Gamma=0}_{11}$. Using this relationship between $S'_{12}$ and
$S'_{11}$ in Eq. (\ref{eq:SS'}) we obtain
\begin{equation}
S_{11} = S'_{11} + 
\frac{1}{2}\left(\frac{\nu_1}{\nu_2}-1\right)S'^{\Gamma=0}_{11} \ 
\end{equation}
which is equivalent, upon separating decoupled and impurity components of the
noise, to
\begin{eqnarray}
S^{\Gamma=0}_{11} &=& 
\frac{1}{2}\left(\frac{\nu_1}{\nu_2}+1\right) S'^{\Gamma=0}_{11} \nonumber \\
&=&\frac{1}{2}\left(\frac{\nu_1}{\nu_2}+1\right) 
g'  {\frac{e^2}{2\pi}} |\omega| \coth{\frac{|\omega|}{2T}}\nonumber \\
&=& \nu_1 {\frac{e^2}{2\pi}} |\omega|  \coth{\frac{|\omega|}{2T}}
\label{eq:decoupled-noise}
\end{eqnarray}
and
\begin{equation}
S^{\rm imp}_{11} = S'^{\rm imp}_{11}\ 
\label{eq:coupled-noise} 
\end{equation}

It is important to observe this distinct behavior for the decoupled
and impurity components of the noise. The important result to be
extracted from Eqs.~(\ref{eq:decoupled-noise}-\ref{eq:coupled-noise})
is that
\begin{enumerate}
\item
The noise in the current of each branch, in the $\Gamma=0$ limit,
depends only on the filling fractions $\nu_1,\nu_2$ on either side of
the junction, and as such it scales with the conductances.
\item
The impurity contribution to the noise $S^{\rm imp}$ 
is exactly the same as for the rotated basis, with an
effective $g'$ given by the harmonic average of $\nu_1$ and
$\nu_2$. $S^{\rm imp}$  depends on the {\it combined} properties of the 
{\it coupled system}.
\end{enumerate}

Let us consider now the behavior of the noise in the different
branches.

\subsection{Noise in the Incoming Channels}

We will focus primarily on the auto-correlations of the densities on
the FQH and the Fermi Liquid (FL) sides of the junction, so it is
useful to introduce an explicit notation
\[
S^{\rm \ in}_{\rm FQH}=S^{\rm in}_{11} \qquad {\rm and} \qquad
S^{\rm in}_{\rm FL}=S^{\rm in}_{22}
\]

More explicitly, these correlations can be written in terms of the
densities and the definitions in
Eqs. (\ref{eq:sijofomega},\ref{eq:def-in}): given by:
\begin{eqnarray}
S^{\rm \ in}_{\rm FQH} &=&  \int_{-\infty}^{\infty} dt\; \cos\omega t\;
\langle \{\rho_1(x<0,t), \rho_1(x<0, 0)\} \rangle \nonumber \\
S^{\rm in}_{\rm FL}&=&  \int_{-\infty}^{\infty} dt\; \cos\omega t\;
\langle \{\rho_2(x<0,t), \rho_2(x<0, 0)\} \rangle \nonumber \\
&&
\label{eq:incom}
\end{eqnarray}

Using Eqs.(\ref{eq:rot1}) and (\ref{eq:rot3}), the densities
$\rho_{1,2}$ can be written in terms of the densities of the fields
${\tilde \phi}_{+,-}$:
\begin{eqnarray}
\rho_1(x<0, t) &=& \sqrt{\frac{g'\nu}{2}} {\tilde \rho}_+(x<0, t) +
\sqrt{\frac{g'}{2}} {\tilde \rho}_-(x<0, t)    \nonumber \\
\rho_2(x<0, t) &=& \sqrt{\frac{g'}{2\nu}} {\tilde \rho}_+(x<0, t) -
\sqrt{\frac{g'}{2}} {\tilde \rho}_-(x<0, t) \nonumber \\
&&
\label{eq:incnoib}
\end{eqnarray}

The noise in the incoming channels is then reduced to a sum of two
terms, one involving the noise in the field ${\tilde \rho}_+$ and the
other involving the noise in the field ${\tilde \rho}_-$. As mentioned
in the previous section, the field ${\tilde \rho}_+$ is
free and its noise is given by Eq.(\ref{eq:sp}). For the noise in the
field ${\tilde \rho}_-$ for $x < 0$ we use Eqs.~(\ref{eq:psixplus}-
\ref{eq:nzT}) to obtain:
\begin{equation}
S_-(\omega, x < 0) = {\frac{e^2}{2\pi}} |\omega|\coth{\frac{|\omega|}{2T}}
\label{eq:sm}
\end{equation}

Using the fact that $\langle {\tilde \rho}_+ {\tilde \rho}_- \rangle = 0$ and
Eqs.(\ref{eq:sp}, \ref{eq:sm}) we find:
\begin{eqnarray}
S^{\rm \ in}_{\rm FQH} &=& \nu {\frac{e^2}{2\pi}} |\omega|
\coth{\frac{|\omega|}{2T}} \nonumber \\
S^{\rm in}_{\rm FL}  &=& {\frac{e^2}{2\pi}} |\omega|
\coth{\frac{|\omega|}{2T}}
\label{eq:incnoise}
\end{eqnarray}

These results agree with results obtained in Ref.~\cite{CFW} for the
incoming branches as expected. The incoming branches are insensitive
to the tunneling between the edges as a result of their chirality.

\subsection{Noise in the Outgoing Channels}

Again, we will focus primarily on the auto-correlations of the
densities on the FQH and the Fermi Liquid (FL) sides of the junction,
and so we introduce an explicit notation
\[
S^{\rm \ out}_{\rm FQH}=S^{\rm out}_{11} \qquad {\rm and} \qquad
S^{\rm out}_{\rm FL}=S^{\rm out}_{22}
\]

The expressions for the noise in the outgoing channels are given by:
\begin{eqnarray}
S^{\rm \ out}_{\rm FQH}&=&  \int_{-\infty}^{\infty} dt\;\cos\omega t\;
\langle \{\rho_1(x>0,t), \rho_1(x>0, 0)\} \rangle \nonumber \\
S^{\rm out}_{\rm FL}   &=&   \int_{-\infty}^{\infty} dt\;\cos\omega t\;
\langle \{\rho_2(x>0,t), \rho_2(x>0, 0)\} \rangle \nonumber \\
& & 
\label{eq:out}
\end{eqnarray}

In order to write the densities $\rho_1(x>0,t)$ and $\rho_2(x>0,t)$ in
terms of ${\tilde \rho}_{+,-}$, we use Eqs.(\ref{eq:rot3}) and
(\ref{eq:rot11}):
\begin{eqnarray}
\rho_1(x>0, t) &=& \sqrt{\frac{g'\nu}{2}} {\tilde \rho}_+(x>0, t) -
\sqrt{\frac{g'}{2}} {\tilde \rho}_-(x>0, t)    \nonumber \\
\rho_2(x>0, t) &=& \sqrt{\frac{g'}{2\nu}} {\tilde \rho}_+(x>0, t) +
\sqrt{\frac{g'}{2}} {\tilde \rho}_-(x>0, t) 
\nonumber \\
& &
\label{eq:outnoib}
\end{eqnarray}

As we have done previously when discussing the noise in the current
$I$, we will express the noise in the incoming and outgoing channels
in terms of $T_k$, the temperature $T$, the frequency $\omega$, and
the voltage $V$, through the Josephson frequency $\omega_0=eV$ (in
units with $\hbar=1$).  After some algebra we find,
\begin{eqnarray}
S^{\rm \ out}_{\rm FQH}&=& \nu {\frac{e^2}{2\pi}} |\omega|
\coth{\frac{|\omega|}{2T}}  + 
S^{\rm imp}\left(\omega,V,T,T_k \right) \nonumber \\
S^{\rm out}_{\rm FL}&=& {\frac{e^2 }{2\pi}} |\omega|
\coth{\frac{|\omega|}{2T}} \ +  S^{\rm imp}
\left(\omega,V,T,T_k\right)
\label{eq:outnoise}
\end{eqnarray}

\breakon

where
\begin{eqnarray}
S^{\rm imp} &=& \frac{T_k}{2\pi } \int_{-\infty}^{\infty} dx \left\{ \frac{x (2 x - \omega/T_k)}{(x^2 + 1) [(x - \omega/T_k)^2 + 1]} \left[ n_{T_k x - \omega}\;  n_{-T_k x} + (1 - n_{\omega -T_k x)})(1 - n_{T_k x})\right] + (\omega \to -\omega) \right\} \nonumber \\
&-& \frac{T_k}{2\pi } \int_{-\infty}^{\infty} dx \left\{ \frac{4 x (x - \omega/T_k) + (\omega/T_k)^2}{(x^2 + 1) [(x - \omega/T_k)^2 + 1]} n_{T_k x - \omega} (1 - n_{T_k x}) + (\omega \to -\omega)\right\}
\label{eq:long1b} 
\end{eqnarray}

and
\begin{equation}
n_{T_kx} = \frac{1}{1 + e^{T_k/T(x - eV/2T_k)}}
\end{equation}
\breakoff 

One can show that the expressions in Eq.~(\ref{eq:long1}) for the noise in
the current through the barrier $I$ and Eq.~(\ref{eq:long1b}) coincide
in the limit of $T=0$. For non-zero temperatures, cross correlations
between $in$ and $out$ branches that appear in the expression for the
noise in the current $I$ Eq.~(\ref{eq:Ifrom-}) make the results no
longer equal.


\section{Conclusions}
\label{sec:conclusions}

In this paper we have discussed the noise spectrum of generic tunnel
junctions between FQH systems at inequivalent Laughlin filling
fractions $\nu_1$ and $\nu_2$, and discussed in great detail the
special exactly solvable case of a junction between the $\nu =1/3$
state and a normal metal ({\it i.\ e.\/}, $\nu=1$). The main focus of
this work was to obtain the expression for the DC noise of the current
$I$ through the ($\nu_1,\nu_2$) junction in order to determine the
charge of the carriers. Using the single point contact model developed
in Ref.~\cite{cl+ed,andreev} and a suitable set of transformations, we
mapped the original model into an effective model of a junction
between equivalent FQH states at certain filling fractions. This
effective model possesses an exact duality transformation between weak
and strong tunneling amplitudes.

We found that different equivalent descriptions can be used to picture
the physics of noise in these junctions. In one picture, more
transparent in the weak tunneling or strong backscattering regime, the
shot noise is produced by electrons, which are the natural common
excitations of the decoupled system, tunneling through the junction.
In contrast, in the weak backscattering or strong tunneling limit, we
obtained a quite interesting result.  We found that the level of DC
shot noise can be attributed to the tunneling of a fractionally
charged state whose charge does not coincide with the charges of the
quasiparticles of the isolated Laughlin states on either side of the
junction. Instead, the fractionally charged state governing the
tunneling properties in this regime can be best regarded as a soliton
of the {\it coupled system}. The charge of the solitons is, in units
of the electron charge, the harmonic average of the filling fractions
of the two Laughlin states. Furthermore, as pictured in
Fig.~\ref{fig:fig-nancyparadox}, the same soliton states (and hence
the same conductance) arise in all junctions between Laughlin states
with the same harmonic mean.  Therefore, any two-terminal measurement
of these junctions cannot distinguish one from another and,
consequently, have no means of determining uniquely the quantum
numbers of the quasiparticles of the bulk Laughlin states of the
junction.

The feature that the noise level tracks the saturation differential
conductance is a general property of QPCs between Laughlin
states. Thus, the only way to tell if the shot noise level is a
measurement of the fractional charge of a state or if instead is a
measurement of the differential conductance is to carry out noise
experiments in fractions which are not in the Laughlin sequence. For
example, for $\nu={\frac{2}{5}}$, the charge is ${\frac{1}{5}}$.

The special case of $(1,{\frac{1}{3}})$ can be exactly solved by a
further mapping to an equivalent free fermion system. For this case,
of direct physical interest, we calculated exactly the current
correlation functions of the original junction in terms of the
correlation functions of the fermions. We examined the DC noise in
both the current $I$ and the backscattering current $I_b$ at zero and
finite temperature and for all voltages. We present our results in the
form of a phase diagram in the voltage-temperature plane. This
description treats (non-equilibrium) shot noise and thermal noise on
the same footing. We showed that there is a natural generalization of
Wilson ratios, which we defined as quotients between the shot noise
and the thermal noise levels for both currents $I$ and $I_b$. These
are universal amplitude ratios, they are scaling functions of $V/T$
independent of the coupling constant near the weak and strong coupling
fixed points, and are related by duality. These ratios can be used to
analyze the experimental data in the recent works by L.~Saminadayar
{\it et. al.}
\cite{Glattli} and R.~de Picciotto {\it et. al.} \cite{Reznikov}
for both limits of thermal and shot noise in a unified way.

We also discuss a four probe geometry that allows for the extraction
of more information from noise measurements by looking at the
auto-correlations of the density or current fluctuations on the
incoming and outgoing channels at the junction. In section
\ref{sec:auto-correlations} we discuss a four probe geometry and
calculate the noise in all four channels. As a consequence of
chirality, the noise in the incoming channels is insensitive to the
presence of the QPC, and it is given by the Johnson-Nyquist noise
level, which is proportional to the conductance determined by the
respective filling fraction in either side of the junction. In
contrast, the noise in the outgoing channels contains the information
on the tunneling coupling, and depends on both filling fractions
$\nu_1$ and $\nu_2$.

We finalize by proposing an experimental setup that can test our
results. The geometry that we believe is most promising is depicted in
the Fig.~\ref{fig:fig-hotspot}. This setup is a variant of the
cleaved edge overgrowth used by A.\ Chang and collaborators
\cite{Chang,Matt}. The only (and important) difference is that the
tunneling region where the barrier is significantly lower is
sufficiently narrow to be regarded as a QPC. In practice, for a smooth
barrier, a tunneling region a few cyclotron lengths wide should be
sufficient to produce a few coherent tunneling centers. In fact, many
of the predictions of references \cite{cl+ed,andreev} can also be
tested in this QPCs.

\section{Acknowledgments}

This work has been supported by the National Science Foundation
through the Grant NSF DMR-94-24511, by a fellowship of the University
of Illinois and by the Science and Technology Center for
Superconductivity of the University of Illinois (NPS).  We are
grateful to A. \ Chang, C.\ Glattli, M. \ Grayson and M.\ Reznikov for
many insightful comments and discussions.

\appendix
\section{Calculation of the functions $F_1$ and $F_2$}
\label{sec:app1}

In this appendix we evaluate the integrals $F_1$ and $F_2$ used for
calculating the noise in Sec.~\ref{sec:noiset}. The strategy is to use
the complex plane and several identities involving the $\psi$ function
(logarithmic derivative of the $\Gamma$ function). First we will
evaluate $F_1$ given by
\begin{eqnarray}
F_1 &=& \sinh\left(\frac{V}{2T}\right) \;\;\;\int_{-\infty}^{\infty} dx  \;\;\;f(x) \nonumber \\
f(x) &=&\frac{x^2}{(1 + x^2)^2}\; \frac{\tanh\left(\frac{V}{4T} - \frac{T_k x}{2T}\right)}{\cosh\left(\frac{T_k x}{2T} + \frac{V}{4T}\right) \cosh\left(\frac{T_k x}{2T} - \frac{V}{4T}\right)}
\label{eq:f1ap1}
\end{eqnarray}

To simplify the notation we introduce $a = (1/2) T_k/T$ and $b = (1/4) V/T$.
After a few algebraic manipulations the integral is cast in the form:
\begin{eqnarray}
F_1 &=& \frac{a}{2} \sinh^2(2b) \;\;\;\int_{-\infty}^{\infty} dy \;\;\;f(y) \nonumber \\
f(y) &=&\frac{y^2}{(a^2 + y^2)^2} \;\;\frac{1}{\cosh^2(y + b) \cosh^2(y-b)}
\label{eq:f1ap2}
\end{eqnarray}

It is straightforward to show that $F_1$ can be calculated by complex
variable methods. The integrand has double poles at $z_{\circ} = \pm
ia$ and at $z_n = \pm b + i \pi(n+1/2)$ and $z_n = \pm b - i
\pi(n+1/2)$.  Because the integral is on the real axis from $-\infty$
to $\infty$ it is enough to consider the contribution from the poles
in the upper (or lower) complex plane. After some 
calculations it can be shown that $F_1$ is given by
\begin{eqnarray}
&&F_1 = \frac{\pi}{4} \sinh^2(2b) \frac{1 + 2a [\tan(a-ib) + \tan(a+ib)]}{\cos^2(a+ib) \cos^2(a-ib)} \nonumber \\
&&- \frac{2i\pi a}{\tanh(2b)} \sum_{n = 0}^{\infty} \Biggl\{\frac{[b + i\pi (n+1/2)]^2}{\bigl[a^2 + \bigl(b +i\pi(n+1/2)\bigr)^2\bigr]^2} - (b \rightarrow -b) \Biggr\}  \nonumber \\
&&- 2i \pi a \sum_{n=0}^{\infty} \Biggl\{\frac{\bigl[b + i\pi(n+1/2)\bigr]\bigl[\bigl(b + i\pi(n+1/2)\bigr)^2 - a^2\bigr]}{\bigl[a^2 + \bigl(b + i\pi (n+1/2)\bigr)^2\bigr]^3} \nonumber \\
&& + \; (b \rightarrow -b) \Biggr\}
\label{eq:f1ap3}
\end{eqnarray}

To further simplify this expression we introduce  the function $S(a,b)$ as
\begin{equation}
S(a,b) = \sum_{n=0}^{\infty} \frac{1}{ a^2 +[b + i \pi(n+1/2)]^2}
\label{eq:sab1}
\end{equation}

In terms of $S(a,b)$ the function $F_1$ is given by:
\begin{eqnarray}
&& F_1 = \frac{\pi}{4} \sinh^2(2b) \frac{1 + 2 a [\tan(a+ib) + \tan(a-ib)]}{\cos^2(a+ib) \cos^2(a-ib)} \nonumber \\
&& - \;\frac{2i\pi a}{\tanh(2b)} \Biggl\{ S(a,b) + \frac{a}{2} \frac{\partial}{\partial a}S(a,b) - c.c.\Biggr\} \nonumber \\
&& + \; \pi i a \frac{\partial}{\partial b} \Biggl\{ S(a,b) + \frac{a}{2}  \frac{\partial}{\partial a}S(a,b) -c.c. \Biggr\}
\label{eq:fvsS}
\end{eqnarray}

The $S(a,b)$ function can be related to the $\psi$-function (digamma)
as follows:
\begin{equation}
S(a,b) = \frac{1}{2\pi a} \left[ \psi\left(\frac{1}{2} - \frac{a+ib}{\pi}\right) - \psi\left(\frac{1}{2} + \frac{a-ib}{\pi}\right) \right]
\label{eq:sab2}
\end{equation}

By defining $\pi z = a +ib$ we find that $F_1$ is given by:
\begin{eqnarray}
&& F_1 =
\frac{\pi}{4} \sinh^2(2b) \frac{1 + 2a [\tan(\pi z) + \tan(\pi \bar z)]}{\cos^2(\pi z) \cos^2(\pi \bar z)} \nonumber \\
&& - \frac{i}{2 \tanh(2b)} \Biggl\{ \frac{a}{\pi} \left[ \psi^{'}(1/2 + z) - \psi^{'}(1/2 - z) - (z \rightarrow \bar z) \right] \nonumber \\
&& +  \psi(1/2 + z) + \psi(1/2 - z) - (z \rightarrow \bar z)  \Biggr\}  \nonumber \\ 
&&+ \frac{1}{4\pi} \Biggl\{ \psi^{'}(1/2-z) - \psi^{'}(1/2+z) - (z \rightarrow \bar z)  \nonumber \\
&& - \frac{a}{\pi} \left[ \psi^{''}(1/2 +z) + \psi^{''}(1/2 -z) + (z \rightarrow \bar z) \right] \Biggr\}
\label{eq:f1ap4}
\end{eqnarray}
where $\psi^{'}(z)$ and $\psi^{''}(z)$ are the first and second
derivatives of the $\psi$ function with respect to their arguments,
evaluated at $z$.

The calculation of $F_2$ and $F_3$ follows similar steps.  Thus, for the
particular case of the $(1,\frac{1}{3})$ junction, we find that the scaling
functions $F_1$, $F_2$ and $F_3$, which describe the crossover from
weak to strong coupling, are determined by a single meromorphic
function, the digamma function, and its derivatives, evaluated at
$\left({\frac{T_k}{2T}}\pm i {\frac{eV}{4T}}\right)$. The behavior of these
correlation functions, which describe non-equilibrium properties, is
remarkably reminiscent of the analytic properties of thermodynamic
functions of integrable quantum impurity problems discovered recently
by Fendley et.~al\cite{fls,FS2}.
 
\section{Limiting behavior of $F_2$ at $V = 0$}
\label{sec:app2}

We calculate explicitly the expressions appearing in
Sec.~\ref{sec:noiset} for the function $F_2$ at zero voltage in both
regimes: $T_k \gg T$ and $T_k \ll T$.  The expression for $F_2$ at
finite voltage and temperature is given by:
\begin{equation}
F_2 = i \frac{\partial}{\partial b} \left\{ \psi\left(\frac{1}{2} + \frac{a-ib}{\pi}\right) - \psi\left(\frac{1}{2} + \frac{a+ib}{\pi}\right) \right\}
\label{eq:f2app1}
\end{equation}
where $a = (1/2) T_k/T$ and $b = (1/4) V/T$. After taking the derivatives
\begin{equation}
F_2 = \frac{1}{\pi} \left[ \psi^{'}\left(\frac{1}{2} + \frac{a-ib}{\pi}\right) + \psi^{'}\left(\frac{1}{2} + \frac{a+ib}{\pi}\right) \right]
\label{eq:f2app2}
\end{equation}

At zero voltage $b= 0$ and the expression for $F_2$ reduces to
\begin{equation}
F_2 = \frac{2}{\pi} \psi^{'} \left(\frac{1}{2} + \frac{T_k}{2\pi T}\right)
\label{eq:f2app3}
\end{equation}

For $T_k \ll T$ the leading order in $F_2$ is simply
\begin{equation}
F_2 = \frac{2}{\pi} \psi^{'}\left(\frac{1}{2}\right) = \pi
\label{eq:f2app4}
\end{equation}

For $T_k \gg T$ we need to use some relations among $\psi-$ functions with
different arguments:
\begin{eqnarray}
\psi^{'}\left(\frac{1}{2} + z \right) &=& 4 \psi^{'}(1 + 2z) - \psi^{'}(1 + z) \\
\psi^{'}(1 + z)           &=& \psi^{'}(z) - \frac{1}{z^2}
\end{eqnarray}

Finally, we use the asymptotic expansions for the $\psi$ function for large
values of its argument to get
\begin{equation}
F_2 = \frac{4T}{T_k} -  \frac{4{\pi}^2}{3} \left( \frac{T}{T_k}\right)^3 + {\cal O} \left(\frac{T}{T_k}\right)^4
\end{equation}

\end{multicols}


\begin{thebibliography}{99}

\bibitem{Glattli}L.~Saminadayar, D.~C.~Glattli, Y.~Jin and B.~Etienne,
Phys.\ Rev.\ Lett.\ {\bf 79}, 2526 (1997).

\bibitem{Reznikov}R.~de Picciotto, M.~Reznikov, M.~Heiblum, V.~Umansky,
G.~Bunin and D.~Mahalu, Nature {\bf 389}, 162 (1997).

\bibitem{Chang}A.~M.~Chang, L.~N.~Pfeiffer, and K.~W.~West, Phys.\ Rev.\
Lett.\ {\bf 77}, 2538 (1996).

\bibitem{Matt} M.~Grayson, D.~C.~Tsui, L.~N.~Pfeiffer, K.~W.~West and
A.~M.~Chang, Phys.\ Rev.\ Lett.\ {\bf 80}, 1062 (1998).

\bibitem{K&F-noise}C.~L. Kane, and M.~P.~A.~ Fisher, 
Phys.\ Rev.\ Lett.\ {\bf 72}, 724 (1994).

\bibitem{fls}P.~Fendley, A.~W.~W.~Ludwig and H.~Saleur, 
Phys.\ Rev.\ Lett.\ {\bf 75}, 2196 (1995).

\bibitem{FS2}P.~Fendley and H.~Saleur, Phys.\ Rev.\ B {\bf 54}, 10845 (1996).

\bibitem{LS3}F.~Lesage and H.~Saleur, Nuc. Phys. B {\bf 493}, 613 (1997).

\bibitem{CFW1} C. de C. Chamon, D. E. Freed, and X. G. Wen, Phys.
Rev. B {\bf 51}, 2363 (1995).

\bibitem{CFW}C.~de C.~Chamon, D.~E.~Freed, and X.~G.~Wen, Phys.\ Rev.\ B
{\bf 53} 4033, (1996) and references therein.

\bibitem{KF-dual} C.~L. Kane, and M.~P.~A.~ Fisher, Phys.\ Rev.\ B {\bf
46}, 15233 (1992).

\bibitem{cl+ed}C.~de C.~Chamon and Eduardo Fradkin, 
Phys.\  Rev.\ B {\bf 56}, 2012 (1997).

\bibitem{wen-edge} For a review see, X.\ G.\ Wen, 
Adv.\ in Phys.\ {\bf 44},405 (1995).

\bibitem{andreev}Nancy P.~Sandler, Claudio de C.~Chamon, and Eduardo Fradkin.
To appear in Phys.\ Rev.\ B, May 15 (1998).

\bibitem{XGWcll}X.~G.~Wen, Phys.~ Rev.~B  {\bf 41}, 12838 (1990).

\bibitem{andrei}N.~Andrei, K.~Furuya, and J.~H.~Lowenstein, Rev.\ Mod.\
Phys.\ {\bf 55}, 331 (1983).



\end{thebibliography}
\end{document}